\documentclass[11pt,a4paper]{article}
\usepackage{times,latexsym}
\usepackage{url}
\usepackage[T1]{fontenc}
\usepackage{authblk}

\usepackage[acceptedWithA]{tacl2021v1}
\usepackage{graphicx}
\usepackage{amsmath}
\usepackage{subcaption}
\usepackage{booktabs}
\usepackage{adjustbox}
\usepackage{multirow}
\usepackage{bbm}
\usepackage{xcolor}

\usepackage{xspace,mfirstuc,tabulary}

\newif\iftaclinstructions
\taclinstructionsfalse 
\iftaclinstructions
\renewcommand{\confidential}{}
\renewcommand{\anonsubtext}{(No author info supplied here, for consistency with
TACL-submission anonymization requirements)}
\newcommand{\instr}
\fi

\iftaclpubformat 

\else

\fi

\title{Can Synthetic Query Rewrites Capture User Intent Better than \\Humans in Retrieval-Augmented Generation?}

\author{
  \textbf{JiaYing Zheng}\textsuperscript{1}, 
  \textbf{HaiNan Zhang}\textsuperscript{1 }\thanks{~\textit{Corresponding author.}}, 
  \textbf{Liang Pang}\textsuperscript{2}, 
  \textbf{YongXin Tong}\textsuperscript{3}, 
  \textbf{ZhiMing Zheng}\textsuperscript{1}
  }
\affil[1]{Beijing Advanced Innovation Center for Future Blockchain and Privacy Computing, 
\authorcr School of Artificial Intelligence, Beihang University, China}
\affil[2]{Institute of Computing Technology, Chinese Academy of Sciences, Beijing, China}
\affil[3]{School of Artificial Intelligence, Beihang University, China}
\affil[ ]{\texttt{\{jiayingzheng, zhanghainan\}@buaa.edu.cn}}

\date{}

\begin{document}
\maketitle
\begin{abstract}

Multi-turn RAG systems often face queries with colloquial omissions and ambiguous references, posing significant challenges for effective retrieval and generation. Traditional query rewriting relies on human annotators to clarify queries, but due to limitations in annotators’ expressive ability and depth of understanding, manually rewritten queries often diverge from those needed in real-world RAG systems, resulting in a gap between user intent and system response. We observe that high-quality synthetic queries can better bridge this gap, achieving superior performance in both retrieval and generation compared to human rewrites. This raises an interesting question: \textbf{Can rewriting models trained on synthetic queries better capture user intent than human annotators?}
In this paper, we propose SynRewrite, a synthetic data-driven query rewriting model to generate high-quality synthetic rewrites more aligned with user intent. To construct training data, we prompt GPT-4o with dialogue history, current queries, positive documents, and answers to synthesize high-quality rewrites. A Flan-T5 model is then fine-tuned on this dataset to map dialogue history and queries to synthetic rewrites. Finally, we further enhance the rewriter using the generator's feedback through the DPO algorithm to boost end-task performance. Experiments on TopiOCQA and QRECC datasets show that SynRewrite consistently outperforms human rewrites in both retrieval and generation tasks. Our results demonstrate that synthetic rewrites can serve as a scalable and effective alternative to human annotations.

\end{abstract}

\section{Introduction}

\begin{figure}[!t]
    \centering
    \includegraphics[width=1.0\linewidth]{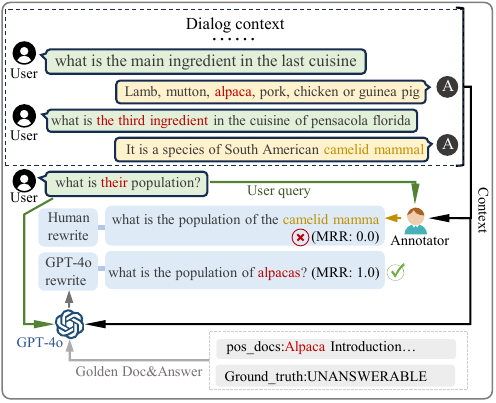}
    \caption{Manual vs. GPT-4o Query Rewriting. }
    \label{fig:compare}
\end{figure}

Retrieval-Augmented Generation (RAG) is widely used in domain-specific tasks like scientific question answering~\cite{che2024hierarchical}, healthcare applications~\cite{zhao2025medrag}, and customer‑service conversational systems~\cite{xu2024retrieval}. In multi-turn dialogues, ambiguous references and colloquial omissions are common in user's current query, requiring a deep understanding of the full dialogue history for accurate intent interpreting~\cite{mao2024rafe,mo2024chiq,fang2022spoken}. Any errors in coreference resolution or omission supplementing can distort query rewrites, harming both retrieval and generation. Thus, the query rewriting task~\cite{ma2023query} is proposed to conduct the complex semantic relationships in dialogue contexts.

Mainstream query rewriting datasets~\cite{anantha2021open,adlakha2022topiocqa} depend on human annotators to clarify user intent from dialogue history, requiring accurate understanding and clear expression. However, due to annotators’ limited expression and understanding, manual query rewrites often fall short of real-world RAG needs, leading to inadequate system responses. For example, in Figure~\ref{fig:compare}, a human annotator incorrectly resolved the pronoun ``their'' in the current query as referring to ``camelid mamma'', whereas the user’s actual intent is ``alpacas''. To quantify this gap, we evaluated retrieval and generation metrics of manually rewritten queries on the TopiOCQA dataset, as shown in Table~\ref{tab:c_train}. Manually rewritten queries yield only 37.7\% MRR@5 and 30.05\% Rouge-L, far below user expectations. Not to mention that models trained on them perform even worse. Therefore, constructing high-quality query rewrites that reflect user intent and suit RAG systems remains challenging.

\begin{table}[!t]
  \centering
  \scriptsize
    \begin{tabular}{lcccc}
    \toprule
    \multicolumn{1}{l}{Method} & \multicolumn{1}{l}{Rouge-1} & \multicolumn{1}{l}{Rouge-L} & \multicolumn{1}{l}{Bleu-4} & \multicolumn{1}{l}{MRR@5} \\
    \midrule
    User Query & -  & - & - & 9.24 \\
    Manual & 33.03 & 30.05 & 26.67 & 37.70 \\ \hline
    Syn\_Unseen & 32.32 & 29.31 & 25.95 & 43.41 \\
    Syn\_Seen & 40.50 & 36.73 & 32.38 & 61.31 \\
    \bottomrule
    \end{tabular}%
  \caption{Comparison of synthetic rewrites and manual rewrites on TopiOCQA training set. Syn\_Unseen generates query rewrites using only the dialogue context and current query, while Syn\_Seen also incorporates the positive document and downstream answer.}
  \label{tab:c_train}%
\end{table}%

To mitigate the gap between query rewriting and downstream components such as the retriever and generator, we employ GPT-4o~\footnote{https://openai.com/research/gpt-4o} to reformulate queries as synthetic data by leveraging positive-relevant documents and gold answers(Syn\_Seen). Based on these synthetic rewrites, we test the retrieval and generation metrics on the TopiOCQA training set, and the results are shown in Table~\ref{tab:c_train}. From the results, we can see that synthetic query rewrites substantially outperform human rewrites in both retrieval and generation tasks. Specifically, in retrieval, it achieves an MRR of 61.31, doubling the performance of human rewrites. In generation, it yields a 23\% improvement in ROUGE-1. Although such results are expected due to the information leakage inherent in the synthetic data construction process(see Section~\ref{sec:leakage}), our primary goal is to investigate \textbf{Whether this synthetic data can be effectively used to train a rewriting model that learns to accurately capture and express the user's true intent}.

In this paper, we introduce SynRewrite, a synthetic data-driven query rewriting model designed to generate queries that more accurately capture the user's underlying intent. Specifically, we leverage ChatGPT-4o to synthesize high-quality rewritten queries, which is prompted with dialogue history, the current user query, relevant positive documents, and their gold answers. Then, we use these synthetic queries to fine-tune a Flan-T5 model, with the dialogue history and current query as input and the synthetic query as the target output. To further improve performance on downstream tasks, we enhance the rewriting model through reinforcement learning using the Direct Preference Optimization (DPO) algorithm, guided by the RAG system's feedback. 

Experiments on two public QA datasets, i.e., TopiOCQA and QRECC, demonstrate that SynRewrite consistently outperforms human annotations in both retrieval and generation tasks.~\footnote{The code and synthetic data are shown in {https://anonymous.4open.science/r/SynRewrite-6C15}.} To assess the generalization capacity of the synthetic rewrites, we also integrate them into the ConvrR model, which consequently achieves superior QA performance compared to the use of human rewrites. That is, the rewriting model can indeed accurately capture the user intent from the synthetic query rewrites, demonstrating their feasibility. 

The innovations of this paper are as follows:
\begin{itemize}
    \item  We explore whether synthetic queries can better capture and express user intent, and validate the feasibility of using synthetic data in query rewriting tasks.

    \item We introduce a query rewriting model based on synthetic data with reinforcement learning, achieving performance that surpasses manually rewritten queries.

    \item Experimental results show that SynRewrite matches human-level performance, and we release our synthetic dataset to support RAG-based query rewriting research.
    
\end{itemize}

\section{Related Work}
In recent years, Retrieval-Augmented Generation (RAG) integrates information retrieval with generation to incorporate dynamic knowledge into LLMs, thereby enhancing question-answering performance~\cite{lewis2020retrieval,zhang2024adacomp}. However, user queries in multi-turn dialogues often involve ambiguity, intent shifts, or context dependence, making query rewriting~\cite{ma2023query} necessary to improve retrieval and generation accuracy~\cite{wang2024depth, owoicho2023exploiting, keyvan2022approach}.

Early query rewriting methods mainly relied on manual rule-based strategies. For instance, CANARD~\cite{elgohary-etal-2019-unpack} addresses coreference and ellipsis in QA tasks by extracting keywords from the context and answer to replace referring expressions. However, such manual rewrites are time-consuming and difficult to scale. With the advent of deep learning, sequence-to-sequence models trained on annotated rewrite datasets~\cite{Svitlana2021, anantha2021open, qian2022explicit, hao2022cgf} can automatically learn rewriting patterns, alleviating the limitations of rule-based methods. ConvGQR~\cite{mo2023convgqr} extends fine-tuned generative models with a knowledge expansion module that integrates semantic rewriting and knowledge injection by generating potential answers and concatenating them with the rewritten query, thereby improving retrieval and downstream generation. Reinforcement learning has also been introduced to further optimize rewriting strategies. CONQRR~\cite{wu2022conqrr} rewrites dialogue questions from the context into standalone questions and employs Self-Critical Sequence Training (SCST)~\cite{rennie2017self} to directly optimize retrieval. MaFeRw~\cite{wang2025maferw} proposes a PPO training framework incorporating multiple feedback signals, further enhancing the performance of rewriting models. Nevertheless, these approaches still rely heavily on high-quality annotations and remain insufficient for handling complex or diverse user intents.

Advances in large language models have further propelled the development of query rewriting techniques. Previous research~\cite{ye2023enhancing} leverages the few-shot response capability of LLMs during the rewriting process, designing prompts to guide LLMs in performing rewriting tasks. 
This approach partially reduces reliance on supervised data. However, in practical applications, it is still necessary to transfer the rewriting ability to smaller models through methods such as distillation, which inevitably incurs performance loss. 
Other studies, IterCQR~\cite{jang2024itercqr} leverage LLMs to synthesize and iteratively refine rewrites with retrieval feedback, though the iterative process introduces additional computational overhead. Furthermore, AdaCQR~\cite{lai2025adacqr} employs a two-stage framework: LLMs generate rewrite labels, followed by Diverse Beam Search to produce and rank candidate queries. While effective, this approach relies heavily on external LLMs and incurs high inference costs.

In summary, existing query rewriting studies in RAG mainly emphasize retrieval gains. This work aims to systematically compare manual and synthetic rewrites across retrieval, generation, offering an analysis of how different rewrite sources capture user intent.

\section{Approach}

\begin{figure}[!t]
    \centering
    \includegraphics[width=1.0\linewidth]{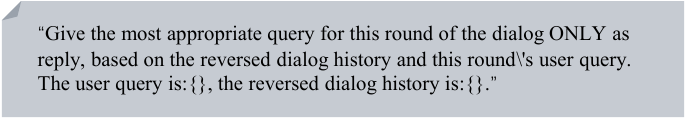}
    \caption{Query Rewriting without gold documents and ground truth. }
    \label{fig:woda promp}
\end{figure}

\begin{figure}[!t]
    \centering
    \includegraphics[width=1.0\linewidth]{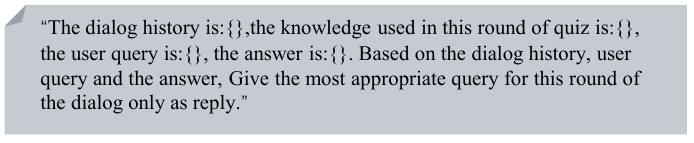}
    \caption{Query Rewriting with gold documents and ground truth. }
    \label{fig:da promp}
\end{figure}
In this section, we will introduce specific methods for training rewriting models. Section~\ref{sec:synthetic} describes the annotation process for synthetic query rewrites, provides a detailed analysis of the synthetic dataset, and examines the advantages of synthetic rewriting over human-generated rewrites. Section~\ref{sec:rewrite} explains the training principles of SynRewrite.
\subsection{Synthetic Query Rewrites}\label{sec:synthetic}
Conventional query rewriting datasets, such as TopiOCQA~\cite{anantha2021open} and QRECC~\cite{adlakha2022topiocqa}, rely extensively on manual annotation. Human annotators are required to read the dialogue history together with the user’s current query, and then rewrite the query into a complete and unambiguous form for retrieval, especially in cases involving missing entities or ambiguous references. This process is not only costly but also difficult to ensure consistent data quality. Moreover, prior works~\cite{ye2023enhancing,jang2024itercqr} have also demonstrated that human rewrites are not always the optimal reformulations. With the rapid progress of large language models (LLMs), we propose to replace human annotators with state-of-the-art LLMs for query rewriting. Our objective is to explore how to leverage LLMs to better capture users’ underlying intentions and to develop more effective methods for query reformulation.

\subsubsection{Data Conduction}
Formally, let the user’s query at turn $n$ be denoted as $q_n$. The dialogue history is defined as $H_n=\{(q_{n-1},a_{n-1}),(q_{n-2},a_{n-2}),…,(q_0,a_0)\}$, the corresponding relevant document is denoted as $pos\_doc_n$, and the gold-standard answer is denoted as $a_n$. We construct synthetic data under two different conditions:

1.	\textbf{Syn\_Unseen.} The LLM receives only the dialogue history $H_n$ and the current user query $q_n$, analogous to the setting of human annotators. It is tasked with generating the most appropriate reformulated query. The corresponding prompt is shown in Figure~\ref{fig:woda promp}.

2.	\textbf{Syn\_Seen.} In addition to the dialogue history $H_n$ and current query $q_n$, the LLM is also provided with $pos\_doc_n$ and $a_n$. In this setting, the model performs rewriting with knowledge of the retrieval target and the gold-standard answer. The corresponding prompt is shown in Figure~\ref{fig:da promp}.

In methods such as AdaCQR, the first-stage soft labels are generated without leveraging extra gold documents or reference answers. Therefore, we investigate two annotation setups—SynSeen and SynUnseen—to examine how additional contextual information affects the quality of synthetic query rewrites, in terms of both retrieval performance and downstream generation accuracy.

\begin{figure}
    \centering
    \includegraphics[width=1.0\linewidth]{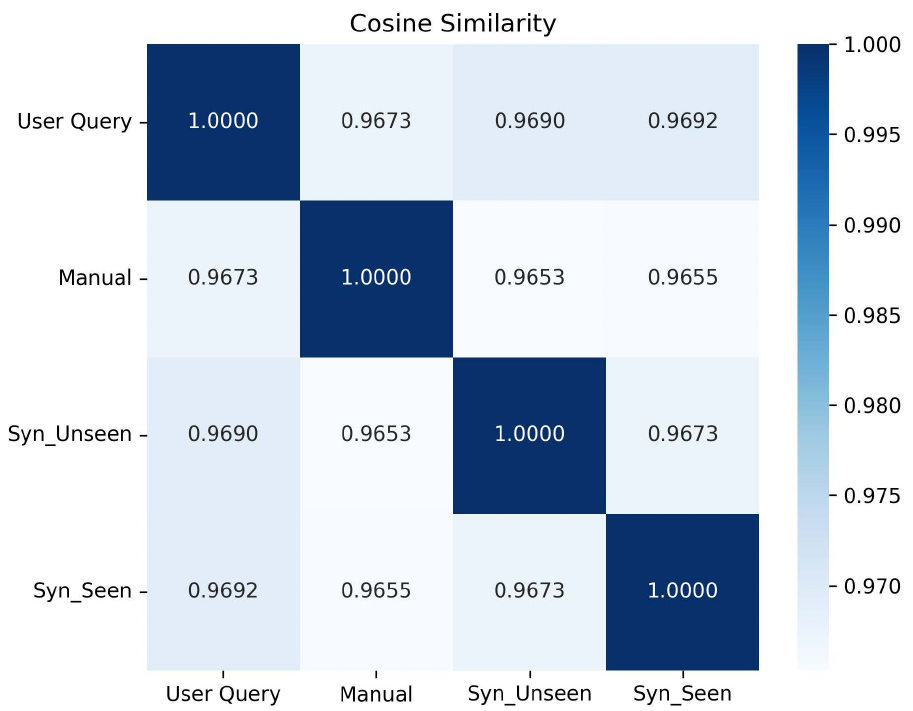}
    \caption{Cosine similarity of different rewrite queries.}
    \label{fig:cos}
\end{figure}

\subsubsection{Data Analysis} 

First, we conduct an analytical evaluation of the training datasets generated by different rewriting strategies, testing both generation metrics and retrieval metrics to clarify the specific performance of each rewriting approach. We denote Syn\_Unseen as the setting where GPT-4o performs rewriting without access to the gold retrieval documents and reference answers, and Syn\_Seen as the setting where GPT-4o leverages the gold retrieval documents and reference answers during rewriting. It is worth noting that, in this section, the generation metrics are still computed by feeding the rewritten queries together with the gold retrieval documents into the generation module.

For each original user query, four types of query variants are available: user query, manual, Syn\_Unseen, and Syn\_Seen. We compute their statistical characteristics in terms of query length, and the mean token counts are 6.91, 8.78, 8.50, and 8.51, respectively. This shows that the three rewritten forms are on average about 25\% longer (in token count) than the original queries, indicating that the rewriting process indeed expands and enriches the content.

To further analyze the semantics of the rewrites, we sample 1,000 instances and encode the four query types using an embedding model. By computing cosine similarity among their semantic vectors based on msmarco-roberta-base-ance-firstp embedding model~\footnote{https://huggingface.co/sentence-transformers/msmarco-roberta-base-ance-firstp}, we obtain the heatmap shown in Figure~\ref{fig:cos}. The average semantic similarity (cosine similarity) across different rewriting methods lies between 0.965 and 0.969, suggesting that all three rewriting methods are semantically very close to the original queries in embedding space, i.e., the rewriting process does not deviate from the original user intent. Specifically, the similarity between human rewrites and original queries is slightly lower than that of LLM-based rewrites, while the Syn\_Seen rewrite is even closer to the original query than human rewrites and Syn\_Unseen rewrites.


Table~\ref{tab:c_train} summarizes the generation and retrieval metrics of the three rewrite types, and compares with the original user queries(RAW). GPT-4o rewrites with access to gold retrieval documents and reference answers substantially outperform human rewrites, with gains up to 62.6\%. Without gold documents and answers, it achieves high retrieval effectiveness (MRR) but lower generation metrics than manual rewrites, which can be attributed to the lack of gold document and answer information, limiting the model’s ability to produce query rewrites that preserve detailed answer content. The Syn\_Seen rewrite consistently achieves the best results, showing that incorporating both document and answer information enhances retrieval-friendliness and generation accuracy, producing higher-quality training data.

\begin{table}[!t]
  \centering
  \footnotesize
    \begin{tabular}{lcc}
    \toprule
    Method & \multicolumn{1}{l}{Avg\_LR} & \multicolumn{1}{l}{Avg\_PureLR} \\
    \midrule
    Manual & 0.0582 & 0.0618 \\
    Syn\_Unseen & 0.0601 & 0.0641 \\
    Syn\_Seen & 0.1015 & 0.1081 \\
    \midrule
    ConvGQR & 0.0529 & 0.0552 \\
    SynConvGQR(Syn\_Seen) & 0.0615 & 0.0652 \\ \hline
    SynRewrite(Syn\_Seen) & 0.0524 & 0.0559 \\
     \ \ \ w.o RL & 0.0491 & 0.0531 \\

    \bottomrule
    \end{tabular}
  \caption{Average and pure entity leakage ratios for various query rewriting approaches. “Syn-” indicates training on synthetic datasets, with the content in parentheses specifying the type of synthetic data. w.o RL denotes fine-tuning without reinforcement learning}
  \label{tab:leak}%
\end{table}%

\subsubsection{Why Synthetic Rewrites Better?} \label{sec:leakage}


By comparing the metrics discussed above, it is evident that the data obtained through rewriting with access to gold retrieval documents and reference answers exhibits a clear advantage. The inclusion of such auxiliary information significantly improves rewriting accuracy. This improvement, however, can be reasonably attributed to unavoidable answer leakage introduced by the additional data. To quantify this effect, we evaluate the proportion of entities in the rewritten queries that are absent from the dialogue history but originate from the positive documents (posdoc) or the ground truth answers.

Concretely, we employ the standard small spaCy model “en\_core\_web\_sm”~\footnote{https://huggingface.co/spacy/en\_core\_web\_sm} to extract entities from text. Let the entity sets extracted from the rewritten query, the gold retrieval documents and answers, and the dialogue history be denoted as Ent(query), Ent(docans) and Ent(history), respectively.
\begin{itemize}
    \item Let N denote the number of entities in Ent(query).
    \item Let M denote the number of entities in Ent(query) that do \textbf{not} overlap with Ent(history).
    \item Let K denote the number of entities in Ent(query) that originate \textbf{solely} from Ent(docans).
\end{itemize}
We define the leakage rate as the proportion of query entities that originate solely from the gold documents and answers:
\begin{align}
    LR&=K/ M, \{0, if M=0\}.
\end{align}
The pure leakage rate is defined as the proportion of query entities that do not belong to Ent(history) but originate from Ent(docans):
\begin{align}
    PureLR&=K/N, \{0, if N=0\}.
\end{align}
Finally, calculate the average across all samples, denoted as Avg\_LR and Avg\_pureLR.


As shown in Table~\ref{tab:leak}, experimental results reveal a clear trade-off between query rewriting effectiveness and sensitive entity leakage risk. Specifically, while GPT-4o demonstrates robust rewriting capabilities, it also exhibits the highest leakage rates, indicating a near-linear correlation between information richness and exposure risk.  Notably, human rewriting also produces non-negligible leakage (about 6\%), indicating that entity exposure is unavoidable. Compared to synthetic methods, rewrite models which trained on synthetic data,  SynRewrite and SynConvGQR maintain controllable leakage levels while providing moderate rewriting improvements.

\subsection{Rewriting Model}\label{sec:rewrite}
This section employs first supervised fine-tuned(SFT), then undergoes reinforcement training(RL) to enable the base model to fully learn rewriting preferences, yielding query rewrites that better align with specific evaluation criteria.
\subsubsection{SFT stage}
We conducted supervised fine-tuning (SFT) on the pre-trained Flan\_t5\_large model~\cite{longpre2023flan} using the three types of rewritten data mentioned above. The training objective is to minimize the standard autoregressive cross-entropy loss:

\begin{equation}
    \mathcal{L}_{\text{CE}}(\theta) = - \frac{1}{N} \sum_{i=1}^{N} \sum_{t=1}^{T_i} \log \pi_\theta \big(y_{i,t} \mid y_{i,<t}, x_i \big).
\end{equation}
Here, $x_i$ denotes the input sequence (including historical dialogues and user questions, $y_i = (y_{i,1}, \dots, y_{i,T})$ represents the corresponding rewriting sequence, and $\pi_\theta$ is the parameterized generative model distribution.

\subsubsection{RL stage}
After the first-stage fine-tuning, the model has acquired basic rewriting ability. In the next step, we aim to further improve the performance of the rewriting model using Direct Preference Optimization (DPO)~\cite{rafailov2023direct}. The core idea of DPO is to eliminate the dependence of RL on explicit reward models by modeling preference data through the Bradley-Terry framework. Specifically, each human preference sample consists of a positive response (“chosen”) and a negative response (“rejected”), where the chosen response is preferred over the rejected one. By quantifying such pairwise preferences, DPO guides the policy model to produce outputs closer to the chosen responses. Formally, given a preference triplet $(x,y^+,y^-)$, the optimization objective of DPO is defined as:

\begin{equation}
\begin{split}
\mathcal{L}_{DPO}(\theta)= & \mathbbm{E}[log\sigma(\beta((\Delta \theta^+ -\Delta \theta^- ) \\ 
&-(\Delta\text{ref}^+ -\Delta\text{ref}^- )))].
\end{split}
\end{equation}

Here, $\Delta_{\theta}^{\pm} = \log \pi_{\theta}(y^{\pm} \mid x)$, $\Delta_{\text{ref}}^{\pm} = \log \pi_{\text{ref}}(y^{\pm} \mid x)$, $\pi_{\theta}$ denotes the policy model to be optimized, while $\pi_{\text{ref}}$ is the reference model. $x$ denotes the input,$y^{+}$ and $y^{-}$ represent the preferred (chosen) and less-preferred (rejected) candidate outputs, respectively. The hyperparameter $\beta$ controls the temperature of preference signal.

However, DPO is highly dependent on the distribution of the preference dataset. Preference pairs with strong contrast make the optimization objective of DPO clearer, but in practice, the contrast in human preference data is often heterogeneous. To address this limitation, the recently proposed Anchor Preference Optimization (APO) \cite{d2025anchored} extends DPO by introducing an additional anchor distribution $\pi_{\text{anc}}$. By regularizing the KL divergence between the policy model and the anchor distribution, APO enables the alignment objective to adapt more flexibly to both the model and the data. The anchor distribution $\pi_{\text{anc}}$ serves as a stable alignment target during optimization, preventing the policy model from relying excessively on the reference distribution $\pi_{\text{ref}}$, thereby theoretically ensuring better generalization and more stable convergence. Its loss function is formally defined as

\begin{equation}
\begin{split}
    \mathcal{L}_{APO} (\theta)= & \mathbbm{E}[log\sigma(\beta((\Delta\theta^+ -\Delta\theta^- )- \\ & (\Delta\text{anc}^+ -\Delta\text{anc}^- )))],
\end{split}
\end{equation}

where $\Delta\text{anc}^\pm =log\pi_\text{anc} (y^\pm\mid x)$, 

APO\_zero is a simplified variant of APO. If it's could be assumed that the chosen outputs are consistently superior to the default outputs of the policy model, then the auxiliary distribution $\pi_\text{anc} $ can be fixed to the initial policy distribution $\pi_\text{*} $. This variant is referred to as APO-zero, whose optimization objective is defined as

\begin{equation}
\begin{split}
  & \mathcal{L}{\text{APO\_zero}}(\theta) = \mathbbm{E}{(x, y^+, y^-)} \\ 
  & \Big[ \big( 1 - \sigma \big( \beta \cdot \text{logratio}_{\text{c}} \big) \big) + \sigma \big( \beta \cdot \text{logratio}_{\text{r}} \big) \Big],
\end{split}
\end{equation}

\begin{align}
\text{logratio}_{\text{c}} = \log \frac{\pi\theta(y^+ \mid x)}{\pi_{\text{*}}(y^+ \mid x)}, \\
\text{logratio}_{\text{r}} = \log \frac{\pi\theta(y^- \mid x)}{\pi_{\text{*}}(y^- \mid x)}.
\end{align}

\section{Experiments}
To demonstrate the performance of models training on synthetic dataset, we define challenge-oriented test sets across four datasets and conduct comparisons with baselines.
\subsection{Experiment Setup}
\subsubsection{Datasets}
We use the training set of the TopiOCQA dataset~\cite{anantha2021open} for annotating synthetic data and evaluating our model performance on its test set. TopiOCQA is an open-domain conversational question answering dataset, consisting of 3,920 dialogues with an average of 13 turns each. The training and test sets contain 45450 and 2514 samples, respectively. It is worth noting that, during the synthetic data annotation stage, GPT-4o leverages the dialogue history, current query, positive documents, and downstream answers as input to more accurately describe the user’s true reformulated query. In the model training and testing stages, however, the input to SynRewrite consists only of the dialogue history and the current query, with the output being the synthetic reformulation. Therefore, there is no risk of answer leakage in our training process, as further demonstrated in Table~\ref{tab:leak}. Additionally, to evaluate the cross-dataset generalization ability of our approach, we also directly test the trained model on the QRECC test set for multi-turn RAG, containing 16,451 samples.

\begin{table*}[!t]
  \centering
  \footnotesize
    \begin{tabular}{clcccccc}
    \toprule
    \multicolumn{1}{l}{Dataset} & Method & \multicolumn{1}{l}{Rouge-1} & \multicolumn{1}{l}{Rouge-2} & \multicolumn{1}{l}{Rouge-L} & \multicolumn{1}{l}{Bleu-4} & \multicolumn{1}{l}{MRR@5} & \multicolumn{1}{l}{EM} \\
    \midrule
    \multirow{3}[2]{*}{\_} & Flan\_t5\_large & 24.12 & 11.12 & 21.94 & 19.23 & 24.60 & 4.49 \\
          & Manual & 32.78 & 17.36 & 29.73 & 26.34 & 53.93 & 4.73 \\
          & GPT-4o & 41.19 & 24.20 & 37.20 & 32.84 & 73.09 & 4.14 \\
    \midrule
    \multirow{2}[2]{*}{Manual} & Flan\_t5\_large & 30.48 & 15.71 & 27.67 & 24.63 & 45.73 & 4.85 \\
          & ConvGQR & 32.55 & 16.98 & 29.45 & 25.93 & 55.25 & 4.61 \\
    \midrule
    \multirow{4}[1]{*}{Synthetic} & Flan\_t5\_large  & 31.51 & 16.97 & 28.69 & 25.57 & 54.14 & 5.09 \\
          & RL\_base & 31.24 & 16.80 & 28.60 & 25.41 & 53.37 & 5.25 \\
          & SynRewrite & 32.79 & 17.44 & 29.83 & 26.56 & 56.34 & 5.37 \\
          & SynConvGQR & 37.70 & 21.02 & 34.17 & 30.35 & 68.39 & 4.89 \\
    \bottomrule
    \end{tabular}%
  \caption{Comparison of generation and retrieval metrics on the TopiOCQA test set.}
  \label{tab:g_r}%
\end{table*}%

\subsubsection{Baselines and Metrics}

We adopt a set of baseline models consisting of three categories: 
\begin{itemize}

    \item \textbf{Flan\_T5\_Large}~\cite{longpre2023flan}. An open-source encoder–decoder model with 780M parameters, derived from T5-large and further instruction-tuned on $\sim$1,000 tasks. It supports inputs up to 1024 tokens and a hidden layer size of 1024. Exhibits strong versatility and instruction-following ability across NLP tasks.

    \item \textbf{RL\_base}~\cite{ma2023query}. To address the issue of user queries in RAG systems being unsuitable for retrieval, a query rewriting task is proposed. First, an LLM generates the query, then a web search engine retrieves the context. Subsequently, a small language model serves as a trainable rewriter, trained via the PPO algorithm using feedback generated by the LLM. It should be noted that due to limited device memory, this method employs offline training.

    \item \textbf{ConvGQR}~\cite{mo2023convgqr}. It is a dialogue query reformulation framework built upon generative pre-trained models on the Manual dataset. Its central idea lies in combining semantic rewriting with knowledge expansion, while incorporating retrieval-oriented signals into the generation process through a knowledge injection mechanism. The framework consists of two independent generative modules: one dedicated to context-aware query rewriting and the other to potential answer generation. The outputs of these two modules are then concatenated to form a decontextualized query suitable for downstream retrieval. It should be noted that the original ConvGQR framework employs T5-base as its backbone; for fairness, all ConvGQR models in this work are built upon Flan\_t5\_large. Since the hidden sizes of the two models differ, we add a linear projection layer to Flan\_t5\_large to align the dimensionality. To assess the generalization capacity of the synthetic rewrites, we also integrate them into the ConvGQR model, denoted as \textbf{SynConvGQR}. 
\end{itemize}

For evaluation, we assess the retrieval model using the performance metric MRR \cite{mrr}, which evaluates how well the system ranks relevant items at the top of the list. And the generation model is evaluated with the metrics ROUGE-1, ROUGE-2, ROUGE-L, \cite{rouge}, BLEU-4 \cite{bleu}, and the QA performance is measured using EM~\cite{zhu-etal-2024-atm}. ROUGE emphasizes recall and n-gram overlap, BLEU balances precision with brevity, and EM measures the exact match rate of generation and gold answer.

\begin{table*}[!t]
  \centering
  \footnotesize
    \begin{tabular}{clccccc}
    \toprule
    \multicolumn{1}{l}{Data type} & method & \multicolumn{1}{l}{Rouge-1} & \multicolumn{1}{l}{Rouge-2} & \multicolumn{1}{l}{Rouge-L} & \multicolumn{1}{l}{Bleu-4} & \multicolumn{1}{l}{EM} \\
    \midrule
    \multirow{3}[2]{*}{\_} &  Flan\_t5\_large & 12.86 & 2.72 & 11.48 & 9.92 & 2.35 \\
          & Manual & 24.98 & 11.25 & 22.69 & 20.22 & 4.49  \\
          & GPT-4o & 34.26 & 18.94 & 31.14 & 27.56 & 4.22  \\
    \midrule
    \multirow{2}[2]{*}{Manual} & Flan\_t5\_large & 23.91 & 10.51 & 21.65 & 19.44 & 4.30 \\
          & ConvGQR & 25.44 & 11.71 & 23.10 & 20.54 & 4.46 \\
    \midrule
    \multirow{4}[2]{*}{Synthetic} & Flan\_t5\_large & 29.85 & 14.99 & 27.23 & 24.21 & 4.77  \\
          & RL\_base & 29.81 & 15.10 & 27.19 & 24.20 & 4.65 \\
          & SynRewrite & 29.63 & 14.81 & 27.24 & 24.33 & 5.81 \\
          & SynConvGQR & 30.90 & 15.92 & 28.08 & 24.83 & 4.89 \\
    \bottomrule
    \end{tabular}%
   \caption{Comparison of RAG performance of different query rewrite methods on TopiOCQA test set.}
  \label{tab:rag}%
\end{table*}%

\subsubsection{Parameter Settings}
We use the open source Flan\_t5\_large pre-trained model~\footnote{http://huggingface.co/google/Flan\_t5\_large} as the basic rewriting model. The model has a hidden layer dimension of 1024 and a parameter count of about 780M, and has been fine-tuned with instructions on a large amount of training data to make the model better at understanding and executing natural language instructions. The training is performed on NVIDIA A100 GPU with 40GB memory, in the fine-tuning phase we set the maximum input length to 1024, the maximum output length to 256, the batch size to 1, the learning rate to 1e-4 and use the Adam optimizer to fine-tune the model by 2 epochs, using the cosine learning rate warmup strategy with a warmup ratio of 0.3. In the reinforcement learning phase, we set the training epoch to 4, the batch size to 2, the gradient accumulation to 8, and the learning rate to 1e-5, also using the linear learning rate warmup strategy, with a warmup ratio of 0.1 and a temperature coefficient of 0.3.

\subsection{Main Results}

\subsubsection{Gold Documents}

We compare the performance of rewrite models trained with various methods on human-annotated data and synthetic data against the baselines, from the perspectives of downstream generation and retrieval effectiveness. It should be emphasized that the downstream generation experiments here evaluate only the performance of the rewritten queries in generative tasks. Therefore, we use the gold positive documents (pos\_doc) from the dataset, rather than documents retrieved by the rewritten queries, and construct the downstream input prompts by combining the rewritten queries with these documents. As shown in Table~\ref{tab:g_r}, on the synthetic rewrite dataset, SynRewrite already surpasses human rewrites across all metrics and achieves the highest EM score. ConvGQR reaches an MRR@5 of 68.4. Moreover, comparing the results across different rewrite datasets reveals that models trained on synthetic data consistently yield better performance. For example, Flan\_t5\_large fine-tuned on synthetic rewrites outperforms its counterpart fine-tuned on human rewrites by an average of 7\%, with an 18\% gain in MRR@5. Similarly, ConvGQR trained on synthetic rewrites improves upon its human-rewrite-trained counterpart by an average of 17.1\%, with an impressive 23.8\% gain in MRR@5.

\subsubsection{RAG performance}
To evaluate the effectiveness of our rewriting model and synthetic data in real RAG pipelines, we apply our rewritten queries from SynRewrite on both the retrieval task and the generation task. The evaluation results of the RAG task are presented in Table~\ref{tab:rag}. We employ the msmarco-roberta-base-ance-firstp dense embedding model to encode the passages from a sampled subset of 75,196 texts from the TopiOCQA corpus. These embeddings are indexed using FAISS (IndexFlatIP) to perform efficient inner-product-based retrieval. Given a user query, we retrieve the top-5 most relevant documents according to their similarity scores for subsequent use in the RAG pipeline. We employ ROUGE, BLEU, and Exact Match (EM) as metrics to assess the extent to which different rewrite models and rewriting strategies enhance the performance of the RAG system. The results clearly indicate that rewrite models can effectively improve the overall performance of RAG. Furthermore, models trained on synthetic data exhibit greater improvements in final generation quality compared to those trained on the human rewrite dataset. Among them, SynRewrite achieves the highest EM, while ConvGQR with synthetic rewrites attains superior ROUGE. These findings suggest that synthetic data not only provides sufficient coverage and diversity for training but also contributes to more robust performance gains across different evaluation dimensions.
\begin{table}[!t]
  \centering    
  \small
  \resizebox{0.5\textwidth}{!}{
    \begin{tabular}{lccccc}
    \toprule
    \multicolumn{1}{l}{Method} & \multicolumn{1}{l}{Rouge-1} & \multicolumn{1}{l}{Rouge-2} & \multicolumn{1}{l}{Rouge-L} & \multicolumn{1}{l}{Bleu-4} & \multicolumn{1}{l}{MRR@5} \\
    \midrule
    \multicolumn{1}{l}{Flan\_t5\_large} & 25.68 & 7.12 & 20.85 & 20.35 & 22.71 \\
    SynRewrite & 31.81 & 10.94 & 25.68 & 26.93 & 32.70 \\ \hline
    ConvGQR & 31.06 & 10.70 & 25.61 & 26.47 & 32.20\\
    SynConvGQR & 33.09 & 11.88 & 26.69 & 28.18 & 34.50 \\
    \bottomrule
    \end{tabular}} %
    \caption{Generalizability of query rewriters: evaluation on QRECC using models trained on TopiOCQA.}
  \label{tab:qrecc_r&a}%
\end{table}%

\begin{table}[!t]
  \centering
  \small
  \resizebox{0.5\textwidth}{!}{
    \begin{tabular}{lccccc}
    \toprule
    \multicolumn{1}{l}{Method} & \multicolumn{1}{l}{Rouge-1} & \multicolumn{1}{l}{Rouge-2} & \multicolumn{1}{l}{Rouge-L} & \multicolumn{1}{l}{Bleu-4} & \multicolumn{1}{l}{EM} \\
    \midrule
    \multicolumn{1}{l}{Flan\_t5\_large} & 27.63 & 9.38 & 22.54 & 22.93 & 0.27 \\
    SynRewrite & 35.18 & 15.23 & 28.76 & 30.48 & 0.54 \\ \hline
    ConvGQR & 35.96 & 15.85 & 29.46 & 31.19 & 0.55 \\
    SynConvGQR & 36.59 & 16.35 & 29.93 & 31.77 & 0.56 \\
    \bottomrule
    \end{tabular} }%
  \caption{Generalizability of RAG performance on QRECC using models trained on TopiOCQA.}
  \label{tab:qrecc_rag}%
\end{table}%

\subsection{Analysis}
To further evaluate other aspects of performance in training rewritten datasets, we conduct the following analytical experiments from multiple perspectives: examining generalization capabilities on other datasets, assessing operational efficiency in real-world RAG scenarios, and conducting case studies.

\subsubsection{Generalization}


To validate the generalization performance of the rewriting model, we tested the model trained on the synthetic TopiOCQA dataset against the QRECC dataset.
The results are shown in Table~\ref{tab:qrecc_r&a} and~\ref{tab:qrecc_rag}. Using the metrics of Flan\_t5\_large (without any fine-tuning) on the QRECC test set as the baseline, it can be observed that all models trained with rewriting significantly outperform the baseline model on metrics such as Rouge, BLEU, EM, and MRR@5. They demonstrate significant improvements in generation metrics (Rouge, BLEU) and exact matching (EM), particularly in MRR@5, indicating enhanced retrieval-friendliness. SynConvGQR outperforms CONVGQR trained on human-annotated data on Rouge-1/2/L, BLEU-4, and MRR@5. The same improvement also appears on Flan\_t5\_large. Overall, the experimental results demonstrate that synthetic data holds potential advantages for cross-dataset generalization.


\begin{figure}
    \centering
    \includegraphics[width=0.65\linewidth]{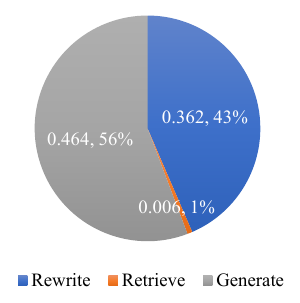}
    \caption{Average processing time(s) of Rewrite, Retrieve, and Generate in RAG.}
    \label{fig:efficiency}
\end{figure}


\begin{figure*}[!t]
    \centering
    \includegraphics[width=0.7\linewidth]{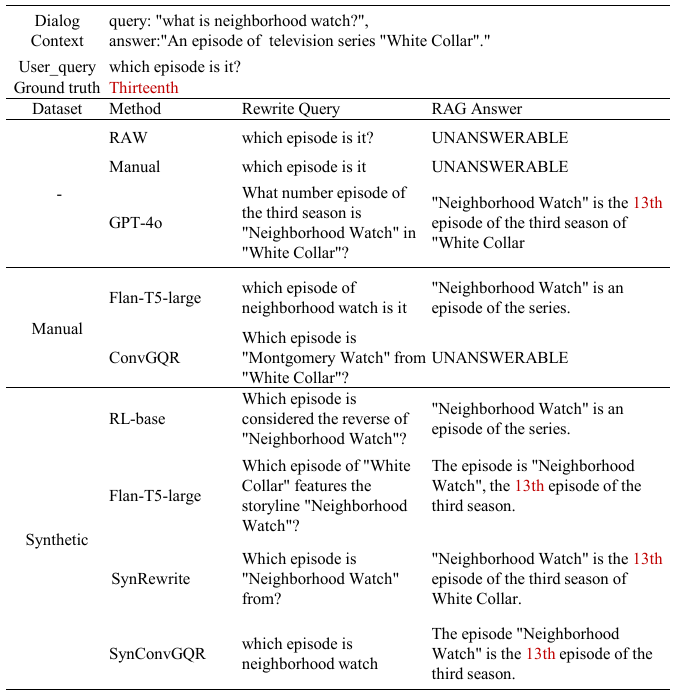}
    \caption{Case study of query rewriting outputs across different rewrite methods. }
    \label{fig:Case study}
\end{figure*}

\subsubsection{Efficiency}
The RAG task places great importance on efficiency, so we tested the rewriting efficiency of the rewriting model with Flan\_t5\_large as the base, as shown in Figure~\ref{fig:efficiency} that in the rag process of 2514 samples from the TopiocQA test set, an average rag takes 0.832 seconds, of which rewriting occupies 0.362 seconds(43\%), retrieval occupies 0.006 seconds(1\%), and generation occupies 0.464 seconds(56\%) , i.e., adding the rewrite model improves the time by 1.77 times than not using the rewrite model. To analyze the performance of the RAG system per unit of time that rewriting can improve, we combine the metrics in Table~\ref{tab:g_r} and~\ref{tab:rag}. Using the SynRewrite rewriting model, the MRR@5 is 56.34 and EM is 5.81; with the SynConvGQR rewriting model, MRR@5 increases to 68.39, but EM drops slightly to 4.89. This indicates that introducing a rewriting module can significantly improve recall (MRR@5), while different rewriting models affect end-answer accuracy (EM) differently.

\subsubsection{Case Studies}
We present a case study from the TopiOCQA test set in Figure~\ref{fig:Case study}, comparing the generation results of various baselines to more intuitively illustrate the differences between human rewriting and synthetic data in RAG tasks. The actual intent of the current query is to identify the episode “Neighborhood Watch” from Season 3 of White Collar. As observed, both the RAW query and the manually rewritten version fail to retrieve or generate results due to missing critical information. In contrast, methods based on synthetic data and those trained on synthetic data consistently incorporate the show title and season number accurately, reliably returning the correct answer. Even when expressed more concisely, the synthetic rewriting model significantly outperforms manually annotated rewrites. This case study validates the advantages of synthetic data and its training strategy in capturing genuine user intent and enhancing RAG generation quality.

\section{Conclusion}
In this work, we examined the limitations of human-annotated query rewrites in retrieval-augmented generation systems and proposed SynRewrite, a synthetic data-driven rewriting framework. By leveraging ChatGPT-4o to synthesize queries guided by dialogue context, relevant documents, and gold answers, SynRewrite provides training signals that better align with real user intent than manual rewrites. Fine-tuning a Flan-T5 model on these synthetic queries, and further optimizing it through reinforcement learning with Direct Preference Optimization allows our approach to significantly improve both retrieval and generation quality.
Experiments on TopiOCQA and QRECC show that SynRewrite surpasses human rewrites and generalizes across architectures like ConvGQR. This demonstrates the effectiveness of synthetic rewrites for bridging user intent and system responses. We release our dataset to spur future work on reducing leakage and extending to broader RAG applications. Looking ahead, we see opportunities in synthetic data construction for users' real intent representation and in extending our method to broader conversational and multimodal RAG applications.

\section*{Limitations}
This study validates the effectiveness of language model-synthesized query rewriting in improving RAG system performance, showing clear advantages over manual rewriting. Experiments primarily relied on GPT-4o-annotated TopiOCQA data, with limited generalization verified on QReCC, leaving performance on larger-scale datasets untested. Future work may explore additional dialogue or cross-domain datasets and methods to enhance practical applicability.


\bibliography{tacl2021v1}
\bibliographystyle{acl_natbib}

\end{document}